\documentclass[aps,amsfonts,nofootinbib,twocolumn]{revtex4}
\usepackage{amsmath}
\usepackage{graphicx}
\usepackage{subfigure}

\newcommand{\beq}{\begin{equation}}
\newcommand{\eeq}{\end{equation}}
\newcommand{\bea}{\begin{eqnarray}}
\newcommand{\eea}{\end{eqnarray}}

\newcommand{\cW}{{\mathcal{W}}}
\newcommand{\dk}{{\int\frac{\d^3 \vc{k}}{(2\pi)^{3/2}}}}

\newcommand{\vc}[1]{{\mathbf{#1}}}

\newcommand{\equa}[1]{\begin{align} #1 \end{align}}

\newcommand{\lh}{\left(}
\newcommand{\rh}{\right)}
\newcommand{\der}{\partial}
\renewcommand{\d}{\mathrm{d}}
\newcommand{\non}{\nonumber}
\DeclareMathSymbol{\mg}{\mathrel}{symbols}{"1D}
\newcommand{\ml}{\ll}

\newcommand{\ga}{\alpha}
\newcommand{\gb}{\beta}
\renewcommand{\gg}{\gamma}
\newcommand{\gd}{\delta}

\newcommand{\gz}{\zeta}

\newcommand{\gth}{\theta}

\newcommand{\gk}{\kappa}

\newcommand{\gf}{\phi}

\newcommand{\gP}{\Pi}

\newcommand{\cH}{{\mathcal H}}

\newcommand{\cJ}{{\mathcal J}}

\newcommand{\cS}{{\mathcal S}}

\newcommand{\tn}{{\tilde n}}

\newcommand{\tge}{{\tilde\epsilon}}

\newcommand{\tgz}{{\tilde\zeta}}
\newcommand{\tget}{{\tilde\eta}}

\begin{document}

\title{Simple route to non-Gaussianity in inflation}

\author{G.I.~Rigopoulos} 
\author{E.P.S.~Shellard}
\author{B.J.W.~van Tent}

\affiliation{Department of Applied Mathematics and Theoretical Physics,\\
Centre for Mathematical Sciences, University of Cambridge,\\
Wilberforce Road, Cambridge, CB3 0WA, United Kingdom}

\begin{abstract}
\noindent We present a simple way to calculate non-Gaussianity in
inflation using fully non-linear equations on long wavelengths with stochastic
sources to take into account the short-wavelength quantum fluctuations.
Our formalism includes both scalar metric and matter perturbations, combining
them into variables which are invariant under changes of time slicing in the 
long-wavelength limit. 
We illustrate this method with a perturbative calculation in the single-field
slow-roll case. We also introduce a convenient choice of variables to
graphically present the full momentum dependence of the three-point correlator.
\end{abstract}

\maketitle

{\it Introduction --- } The standard lore of inflation states that the
statistical properties
of its perturbations correspond to those of Gaussian random fields. Gaussianity
is sometimes presented as a robust prediction regardless of any
realisation in the context of a specific model. Such a statement may be 
adequate for comparing inflation with topological defects, since the
latter are distinctly non-Gaussian, but its validity is
related to the use of linear theory and is only
approximate.\footnote{Gaussianity also depends on
  the definition of the initial state
  of the perturbations \cite{sakell}; we shall assume this to be the
  standard vacuum defined at short wavelengths.} 
Since gravity is inherently non-linear and the
scalar fields responsible for inflation may have interacting
potentials, some non-linearity will always be present and manifest
itself as non-Gaussianity in the cosmic microwave background 
radiation (CMB). Hence the issue is not
whether inflation is non-Gaussian, but how large the non-Gaussianity
is. In an era of precision cosmology, it is potentially
an additional observable to be sought in future CMB missions.

The current observational discriminants between inflationary models 
are the amplitude of the power spectrum, 
the spectral index, its running, and the tensor to scalar ratio \cite{ll}, 
all calculated in linear theory. {\it A priori},
non-Gaussianity offers another quantity, characteristic of a
model, which can be used as a discriminant. Different models are
expected to exhibit a different amount and type of non-linearity and
hence carry their own non-Gaussian signature. Naively, it is expected
that any non-linearity will be very small given that linear
fluctuations are small: an order of magnitude estimate is
\beq
\lh \frac{\Delta T}{T}\rh_\mathrm{NL}\sim f_\mathrm{NL}
\lh \frac{\Delta T}{T}\rh^2_\mathrm{L}.
\eeq
However, given the accuracy of future CMB missions, one could try to
see whether such small non-linearities are observable. For
example, the Planck satellite will be able to detect them, if
$f_\mathrm{NL}\sim\mathcal{O}(1)$ \cite{komatsu}.   

The question of non-Gaussianity requires techniques beyond linear
theory. Apart from early attempts to calculate it
\cite{early_attempts}, the issue attracted some attention recently
after the promise of high quality CMB data from satellites like WMAP
and Planck. The usual theoretical approach has been to expand the Einstein
equations to second order \cite{acquaviva,noh_and_hwang}. An extensive
survey of such techniques (including references) is provided in
\cite{ng_review}. A tree-level action treatment of interacting perturbations 
was given in~\cite{mald}.

In \cite{gp2,formalism} we presented a new framework for calculating
non-Gaussianity. It is valid in a very general multiple-field inflation
setting, which includes the possibility of a non-trivial field metric. 
In this paper we illustrate that general formalism by applying it 
to the simple case of single-field inflation. Explicit calculations in the 
multiple-field case can be found in \cite{mf}.

The formalism consists of a long-wavelength approximation coupled 
with a stochastic picture for the
generation of perturbations on the relevant scales. It describes
the evolution of variables which are invariant under changes of time
slicing in the long-wavelength limit. These variables specify the
inhomogeneous system, taking into
account both matter and scalar metric perturbations, and are
defined without recourse to the concept of a linear
perturbation. The equations are formally very
simple, yet they encode the full non-linear evolution on large
scales. These stochastic equations offer a generalisation of
currently existing stochastic approaches to the description of inflationary 
fluctuations \cite{stoch}. They are suitable for a perturbative
analytic treatment which is technically less complicated than perturbing the
original Einstein equations. They are also well-suited as
a basis for computer simulations which include the full non-linear
evolution of the perturbations. 

This paper is organised as follows. We start by summarising the general 
formalism of \cite{gp2,formalism}, simplified to the single-field case. 
Using this we then analytically calculate the three-point correlator of the
curvature perturbation for
single-field slow-roll inflation, which would be zero in the Gaussian case.
Our result is a completely explicit expression, containing the full momentum
dependence. We introduce a convenient choice of variables to
graphically present this momentum dependence. We also calculate the effects of
the decaying mode.
We conclude by discussing the results.

{\it Basic formalism --- } A defining characteristic of inflation is the
behaviour of the comoving Hubble radius $1/(aH)$, which shrinks
quasi-exponentially. A mode with comoving wavenumber $k$ is called
super-horizon when $k<aH$, and conversely sub-horizon when $k>aH$.  
The inflaton is taken to be in a vacuum state,
defined such that sub-horizon modes approach the Minkowski vacuum for
$k\mg aH$. After a
mode exits the horizon, it is described by a classical probability
distribution with a variance given by the power spectrum \cite{classicality}. 
We will focus attention on the classical super-horizon regime and assume that
the dynamics on these scales are adequately described by dropping from
the equations all terms explicitly containing second-order spatial 
gradients \cite{long wavelength} (i.e.\ the long-wavelength 
approximation).\footnote{Formally this corresponds with taking only the 
leading-order terms in the gradient expansion. We expect higher-order terms 
to be subdominant on long wavelengths during inflation, but this statement has 
only been rigorously verified at the linear level. A calculation to higher 
order in spatial gradients, or, even better, a full proof of convergence of the
expansion, would be desirable.
See \cite{long wavelength, giovannini} for more details on the validity of the
gradient expansion beyond linear theory.}
Then, focusing on scalar modes only, spacetime can be described by the metric
\beq\label{metric}
ds^2=-N^2(t,\mathbf{x})\d t^2+a^2(t,\mathbf{x})\d\mathbf{x}^2,
\eeq     
with $a$ the local scale factor and $N$ the lapse function.
The local expansion rate is defined as 
$H \equiv \dot{a}/(N a)$, 
where the dot denotes a derivative with respect to $t$.
Here, we consider inflation to be driven by a single inflaton 
$\phi$. We also define two {\em local} slow-roll parameters by\footnote{These 
definitions are
equivalent to the more standard ones, $\tge = -\dot{H}/(N H^2)$ and
$\tget = \dot{\gP}/(NH\gP)$, if the standard equations of motion for
$H$ and $\gP$ are used. However, after introducing stochastic source terms (as
below) this relation changes subtly.} 
\beq\label{slowrollpar}
\tge(t,\vc{x}) \equiv \frac{\gk^2 \gP^2}{2 H^2} \,,\qquad
\tget(t,\vc{x}) \equiv -3 - \frac{\der V / \der\gf}{H\gP} \, ,
\eeq
where $V$ is the potential, $\gP\equiv\dot\phi/N$ and 
$\gk^2 \equiv 8\pi G = 8\pi/m^2_{\rm pl}$.
 
The fact that we focus on scalar modes only is an approximation, since beyond 
linear theory, in general, scalar, vector, and tensor modes will mix. For
complete consistency the spatial part of the metric in (\ref{metric}) should be
multiplied by an $h_{ij}$ matrix with unit determinant, containing vector and
tensor modes. However, in the long-wavelength approximation $h_{ij}$ would be 
non-dynamical, as shown for example in \cite{formalism}. Moreover, at second
order in the perturbations (which gives the leading-order non-Gaussianity), the
scalar modes are only affected by the linear vector and tensor modes, the
former of which are zero, while the latter are subdominant to the scalar modes.
Hence it seems a good approximation to neglect vector and tensor modes to obtain
a leading-order expression for non-Gaussianity. Vector and tensor modes will be 
investigated in future work.

Since we will be dealing with non-linearities, it is useful to
describe inhomogeneity without resorting to the concept of a linear
perturbation. In this case,  it makes sense to consider the spatial gradient of
quantities. In particular we will work with the following combination of 
spatial gradients:
\beq\label{gi_var}
\gz_i(t,\vc{x}) \equiv \der_i \ln a(t,\vc{x}) 
- \frac{\gk}{\sqrt{2\tge(t,\vc{x})}} \, \der_i \gf(t,\vc{x}),
\eeq
which is invariant under changes of time slicing, up to second-order
spatial gradients \cite{gp, formalism}. Note that, when linearised,
$\gz_i$ is the 
spatial gradient of the well-known $\gz$ from the literature, the curvature
perturbation.

The full non-linear dynamics of the inhomogeneous system are described entirely
in terms of $\gz_i$. As discussed in \cite{formalism}, one finds a second-order
differential equation for $\gz_i$, with a set of constraint equations expressing
the gradients of the coefficients in this equation in terms of $\gz_i$. These
equations are exact within the long-wavelength approximation.
Introducing the velocity $\gth_i$, we then rewrite the equation of motion as a
system of two first-order differential equations. Finally we introduce
stochastic source terms in both to take into account the continuous flux of
short-wavelength quantum modes crossing the horizon into the long-wavelength
system. We choose the gauge where 
\beq\label{time}
t=\ln(aH) \quad\Leftrightarrow\quad
NH=(1-\tge)^{-1},     
\eeq
since in this gauge horizon exit of a mode, $k=aH$, occurs simultaneously for
all spatial points. Then the full non-linear system of equations for $\gz_i$ 
in single-field inflation can be written as
\cite{formalism}
\beq\label{basic}
\left\{ \begin{array}{l}
\displaystyle
\dot{\gz}_i - \gth_i = \cS_i\\
\displaystyle
\dot{\gth}_i + \frac{3 - 2\tge + 2\tget - 3\tge^2 - 4\tge\tget}{(1-\tge)^2}\, 
\gth_i = \cJ_i
\end{array} \right.
\eeq
where the source terms $\cS_i$ and $\cJ_i$ are defined below in (\ref{sources})
and the constraints are
\bea
\der_i \ln a & = & - \der_i \ln H = - \frac{\tge}{1-\tge} \, \gz_i, \non\\
\der_i \gf & = & - \frac{\sqrt{2\tge}}{\gk} \, \frac{1}{1-\tge} \, \gz_i, 
\label{constr}\\
\der_i \gP & = & - \frac{\sqrt{2\tge}}{\gk} \, H \lh (1-\tge) \gth_i
+ \frac{\tget}{1-\tge} \, \gz_i \rh.
\non
\eea
Since $\tge$, $\tget$, etc.\ depend on $\gz_i$ via the constraints 
(\ref{constr}), the equation of motion (\ref{basic}) is clearly very non-linear.
Note that even though these equations contain slow-roll parameters, 
it is not a slow-roll approximation; the slow-roll parameters are just
short-hand notation for their respective definitions (\ref{slowrollpar}) and 
they have not been assumed to be small.
On the right-hand side of (\ref{basic}), $\cS_i$ and $\cJ_i$ are stochastic 
source terms given by

\bea\label{sourceterms}
\cS_i & = & \frac{-\gk}{2a\sqrt{\tge}} \dk \, \dot{\cW}(k) 
\mathrm{i} k_i \mathrm{e}^{\mathrm{i} \vc{k}\cdot\vc{x}}
Q_\mathrm{lin}(k) \ga(\vc{k}) + \mathrm{c.c.}, \non\\
\cJ_i & = & \frac{-\gk}{2a\sqrt{\tge}} \dk \, \dot{\cW}(k)
\mathrm{i} k_i \mathrm{e}^{\mathrm{i} \vc{k}\cdot\vc{x}}
\label{sources}\\
&& \times \left[ \dot{Q}_\mathrm{lin}(k)
-\frac{1+\tge+\tget}{1-\tge}\,Q_\mathrm{lin}(k) \right] 
\ga(\vc{k}) + \mathrm{c.c.}, \non
\eea
where c.c.\ denotes the complex conjugate. $\cW(k)$ is the Fourier transform of 
an appropriate smoothing window function which cuts off modes with wavelengths 
smaller than the Hubble radius; we choose a Gaussian with smoothing 
length $R\equiv c/(aH) = c \, \mathrm{e}^{-t}$, where \mbox{$c\approx\,$3--5:}
\beq
\cW(k)={\rm e}^{-k^2R^2/2}.
\eeq 
In our gauge $R$ and hence $\cW$ do not depend on $\gz_i$. The perturbation
quantity $Q_\mathrm{lin}$ is the solution from linear theory for the
Sasaki-Mukhanov variable $Q \equiv - a \sqrt{2 \tge} \, \gz / \gk$. 
It can be computed
exactly numerically, or analytically within the slow-roll approximation (see
e.g.\ \cite{mfb, vantent}). 
Finally, $\alpha(\mathbf{k})$ is a Gaussian complex random 
number satisfying
\beq
\langle{\alpha}(\mathbf{k}){\alpha^{*}}(\mathbf{k}')\rangle
=\delta^3\!\left(\mathbf{k}-\mathbf{k'}\right),
\qquad
\langle{\alpha}(\mathbf{k}){\alpha}(\mathbf{k}')\rangle=0.
\eeq

Since $\gz_i$ and $\gth_i$ are smoothed long-wavelength variables, the
appropriate initial conditions are that they should be zero at early times 
when all the modes are sub-horizon. Hence,
\beq
\lim_{t \rightarrow -\infty} \gz_i = 0,
\qquad
\lim_{t \rightarrow -\infty} \gth_i = 0.
\eeq
The full linear solution $Q_\mathrm{lin}$ contains both the growing and
decaying modes. If we neglect the decaying mode (see the end of the next section
for remarks about the validity of this), it was shown in \cite{gp2} that in the
single-field case $\dot{Q}_\mathrm{lin} = NH(1+\tge+\tget)Q_\mathrm{lin}$
(actually this result is true even beyond linear order). Then the system 
(\ref{basic}) simplifies considerably, because this means that 
$\cJ_i = 0$.
With the initial conditions specified above, we then must have that $\gth_i = 0$
at all times. Hence we are left with only
\beq\label{basicsimp}
\dot{\gz}_i = \cS_i.
\eeq
Note that for the fully non-linear case this is still non-trivial to solve,
since $\cS_i$ has a non-linear dependence on $\gz_i$, as can be seen from
(\ref{sources}) in combination with (\ref{constr}). We can either deal
with it directly numerically, or use an approximation method to compute it 
analytically. The numerics are the subject of another paper; here we continue
with the analytic treatment.

{\it Analytic approximations --- } In order to analytically solve
(\ref{basicsimp}) we have to apply two approximations: in the first place 
we only consider
leading-order terms in a slow-roll expansion, and secondly we set up an
expansion in perturbation orders. Hence only in this section do we assume that
$\tge$ and $\tget$ are small; up to now the results were valid for any value.

To leading order in slow roll, $Q_\mathrm{lin}$ is given by (see 
e.g.\ \cite{vantent})
\bea
Q_\mathrm{lin}(k) & = & \sqrt{\frac{\pi}{4k}} \, \sqrt{-k\mathrm{e}^{-t}}
\, \mathrm{H}^{(1)}_{3/2}(-k \mathrm{e}^{-t}) \non\\
& \approx & \frac{\mathrm{i}}{\sqrt{2k}} \lh \frac{1}{k} \, \mathrm{e}^t
- \frac{\mathrm{i}}{3} \, k^2 \mathrm{e}^{-2t} \rh,
\eea
with $\mathrm{H}^{(1)}_{3/2}$ the Hankel function of the first kind and of order
$3/2$. In the second line we have taken the first terms in a series expansion 
for late times; the first term is the growing mode and the second the decaying
mode. We will at first consider only the growing mode, an assumption which 
will be justified at the end of this section. The overall unitary factor is
irrelevant for the correlators and will be omitted.

Equation (\ref{basicsimp}) can now be solved perturbatively. At first order all 
quantities in $\cS_i$ take their homogeneous background values. Rewriting 
$a = c/(RH)$, taking $H$ and $\tge$ to be constant (leading-order slow-roll
approximation), and switching to $R = c\, \mathrm{e}^{-t}$ as integration
variable, the end result is (with $\gz \equiv \der^{-2} \der^i \gz_i$):
\equa{
\gz^{(1)}(\vc{x})
= & -\frac{\gk}{2 \sqrt{2}} \int \frac{\d^3 \vc{k}}{(2\pi)^{3/2}}
\frac{1}{k^{3/2}} \frac{H_\cH}{\sqrt{\tge_\cH}} 
\, \mathrm{e}^{-k^2 R^2 /2} \, \ga(\vc{k}) 
\mathrm{e}^{\mathrm{i} \vc{k}\cdot\vc{x}}
\non\\
& + \mbox{c.c.}
}
Here the subscript $\cH$ denotes evaluation at horizon crossing.
The time-dependent part $\mathrm{e}^{-k^2 R^2 /2}$ actually goes to the constant
value 1 very quickly (in about 3 e-folds after horizon crossing), so that 
$\zeta^{(1)}$ is constant on sufficiently long super-horizon scales, and
independent of the smoothing parameter $c$. Taking $R \rightarrow 0$ we obtain
the standard power spectrum:
\beq
\left\langle |\gz^{(1)}(k)|^2 \right\rangle 
= \frac{\gk^2}{4} \frac{1}{k^3} \frac{H_\cH^2}{\tge_\cH}.
\eeq

At second order in the perturbations we expand all quantities in $\cS_i$ as
follows ($C = a,\tge$, etc.):
\beq
C(t,\vc{x}) = C^{(0)}(t) + C^{(1)}(t,\vc{x})
= C^{(0)} + \der^{-2} \der^i (\der_i C)^{(1)},
\eeq
where we use (\ref{constr}) to compute $\der_i C$. In particular we find
\beq
\der_i \tge = - 2 \tge \lh (1-\tge) \gth_i 
+ \frac{\tge+\tget}{1-\tge} \, \gz_i \rh.
\eeq
Since perturbing $Q_\mathrm{lin}$ only gives next-to-leading-order slow-roll
contributions (a consequence of the fact that $Q$ is the appropriate quantity 
to use on short wavelengths, not $\gz$) and in the gauge (\ref{time}) the 
window function $\cW$ is unperturbed, all non-linearity is actually contained 
in the $a\sqrt{\tge}$ factor in the expression (\ref{sourceterms}) 
for $\cS_i$.
The resulting equation is
\beq\label{second eq}
\dot{\gz}^{(2)}_i = (2\tge^{(0)}+\tget^{(0)}) \, \gz^{(1)} \cS_i^{(1)},
\eeq
with solution 
\bea\label{second sol}
\gz^{(2)}(\vc{x})&=& \frac{\gk^2}{8} \int\!\!\!\!\int \frac{\d^3 \vc{k} \, 
\d^3 \vc{k}'}{(2\pi)^3}
\frac{1}{k^{3/2} {k'}^{3/2}} \frac{H_\cH}{\sqrt{\tge_\cH}}
\frac{H_{\cH'}}{\sqrt{\tge_{\cH'}}} \nonumber\\
&&\times (2\tge_{\cH'}+\tget_{\cH'}) \frac{k^2}{k^2+{k'}^2} \,
\mathrm{e}^{-(k^2+{k'}^2)R^2/2} \nonumber \\
&&\times \Bigg( \frac{k^2 + \vc{k}\cdot\vc{k}'}{|\vc{k}+\vc{k}'|^2} \,
\mathrm{e}^{\mathrm{i} (\vc{k}+\vc{k}')\cdot\vc{x}} 
\ga(\vc{k}) \ga(\vc{k}')\\
&& \quad\; + \frac{k^2 - \vc{k}\cdot\vc{k}'}{|\vc{k}-\vc{k}'|^2} \,
\mathrm{e}^{\mathrm{i} (\vc{k}-\vc{k}')\cdot\vc{x}} 
\ga(\vc{k}) \ga^*(\vc{k}') \Bigg) 
+ \mbox{c.c.}\nonumber
\eea
Again, the time-dependent term $\mathrm{e}^{-(k^2+{k'}^2)R^2/2}$ very quickly
goes to 1, so that $\gz^{(2)}$, like $\gz^{(1)}$, is constant on
sufficiently long super-horizon scales and independent of~$c$.

From (\ref{second sol}) we note that $\langle\zeta^{(2)}\rangle$ is
indeterminate. To remove this ambiguity and also require that
perturbations have a zero average, we define $\tgz \equiv \gz -
\langle \gz \rangle$. Expanding $\tgz=\tgz^{(1)}+\tgz^{(2)}$ and
switching over to Fourier space, we find the three-point correlator 
(or rather, the bispectrum) to be
\bea\label{3pcorr}
&&\left\langle \tgz(\vc{x}_1) \tgz(\vc{x}_2) \tgz(\vc{x}_3)
\right\rangle(\vc{k}_1,\vc{k}_2,\vc{k}_3) \\
&&= (2\pi)^3 \gd^3 \lh
{\textstyle\sum_s} \mathbf{k}_s \rh 
\left [ f(\vc{k}_1,\vc{k}_2) + f(\vc{k}_1,\vc{k}_3) 
+ f(\vc{k}_2,\vc{k}_3) \right ] \nonumber
\eea
with
\bea\label{three point}
f(\vc{k},\vc{k}') & \equiv & 
\frac{\gk^4}{16} \frac{1}{k^3 {k'}^3} \frac{H_\cH^2}{\tge_\cH}
\frac{H_{\cH'}^2}{\tge_{\cH'}} \bigg [ \bigg((2\tge_{\cH'}+\tget_{\cH'})
\frac{k^2}{k^2+{k'}^2} \nonumber\\
&&\times \frac{k^2+\vc{k}\cdot\vc{k}'}{|\vc{k}+\vc{k}'|^2}\bigg) + (\vc{k}
\leftrightarrow \vc{k}') \bigg].
\eea
This result is independent of $c$, so that our choice of smoothing scale does
not matter. Note that it is valid to second order in the perturbations, to 
leading order in slow roll, and on sufficiently long super-horizon scales 
so that $\gz$ has become constant.

In the limit $k_3 \ml k_1,k_2$ (and hence $\vc{k}_1 = - \vc{k}_2
\equiv \vc{k}$), the above expression gives (leaving aside the overall factor 
of $(2\pi)^3 \gd^3(\sum_s \vc{k}_s)$):
\bea
\langle \tgz \tgz \tgz \rangle &=& \frac{\gk^4}{8} \frac{1}{k^3 k_3^3} 
\frac{H_\cH^2}{\tge_\cH}
\frac{H_{\cH_3}^2}{\tge_{\cH_3}} (2\tge_{\cH_3}+\tget_{\cH_3}) \non\\
&=& -\tn(k_3) \langle |\gz(k)|^2 \rangle
\langle |\gz(k_3)|^2 \rangle,
\label{three point 2}
\eea
with $\tn = n$$-$$1$ the scalar spectral index. From (\ref{three point}),
the non-linearity parameter $f_\mathrm{NL}$ has momentum
dependence in general. In the limit where one of the
momenta is much smaller than the other two, $f_\mathrm{NL}\sim \tn$ during
single-field slow-roll inflation. Hence, at least in this limit,
non-Gaussianity is very small in any such inflation model which is
compatible with observations. The result 
(\ref{three point 2}) agrees exactly with the corresponding limit of 
\cite{mald}. A similar conclusion was also reached in \cite{acquaviva}.     

\begin{figure}
\subfigure[]{\label{triangleA}\includegraphics[width=8.6cm]{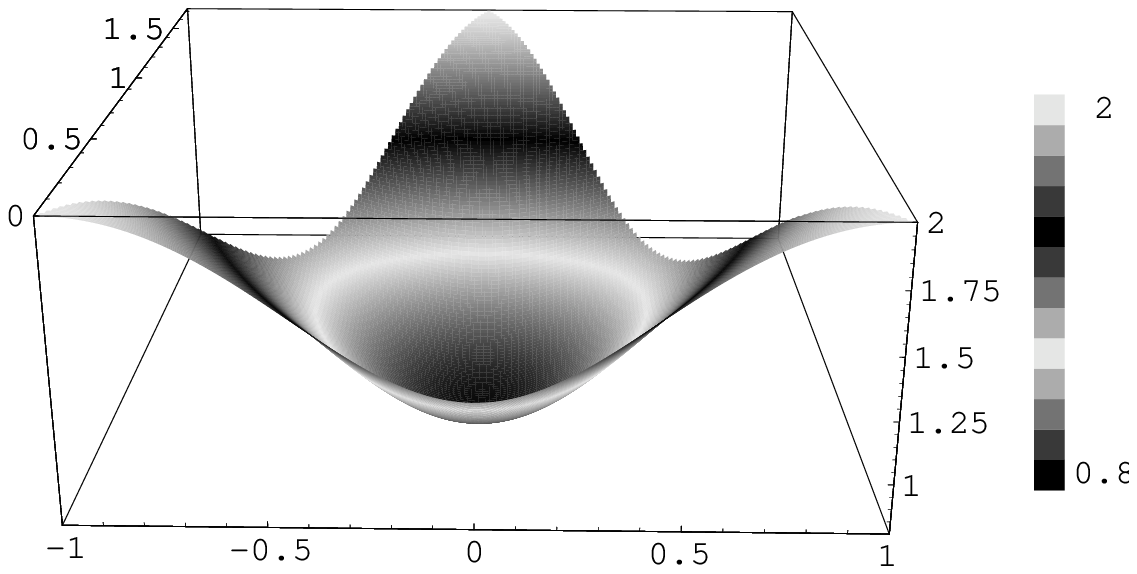}}\\
\subfigure[]{\label{triangleB}\includegraphics[width=5.5cm]{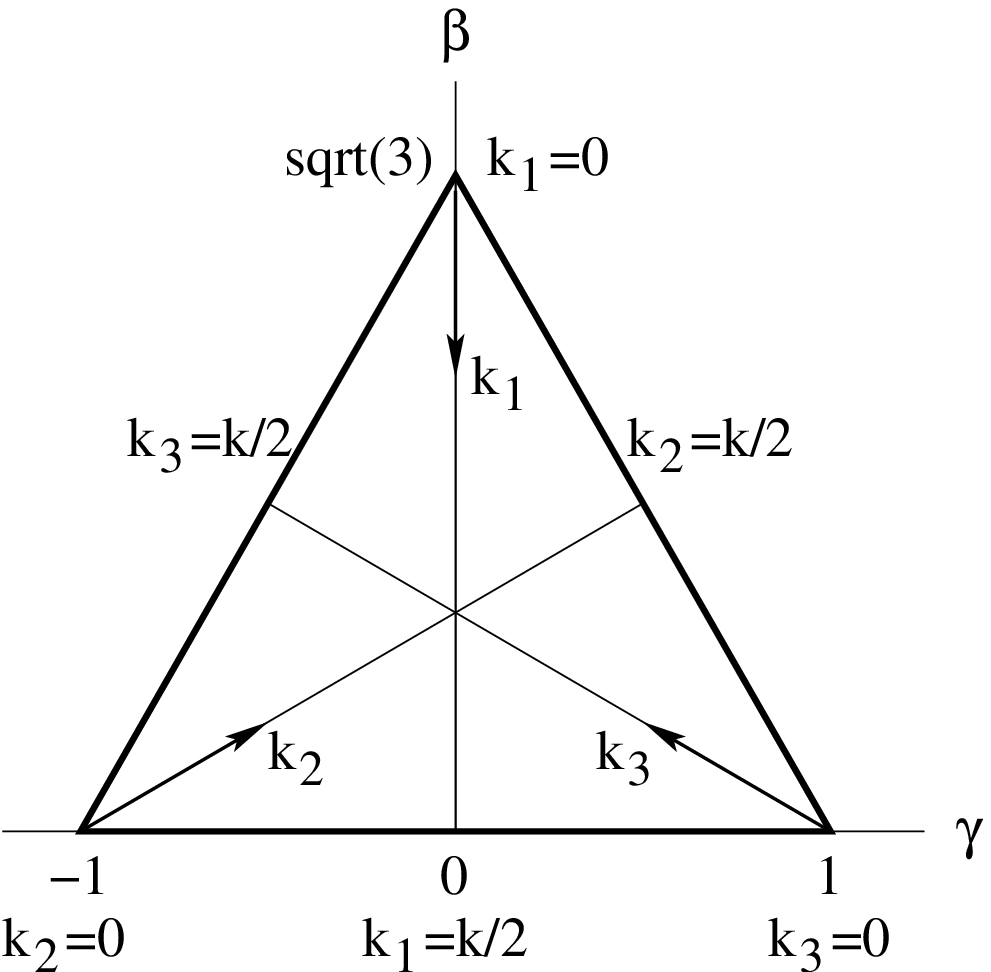}}
\caption{{\em (a)} The three-point correlator 
(\ref{3pcorr}), (\ref{three point}) for single-field slow-roll inflation, 
multiplied by $k_1^3 k_2^3 k_3^3/((k_1^2+k_2^2+k_3^2)/2)^{3/2}
[(\gk^4/16)(H^4/\tge^2)(2\tge+\tget)]^{-1}$, plotted to show its dependence on
the relative size of the three momenta. {\em (b)} An explanation of
the triangular domain used, defined in (\ref{plotvars}).}
\label{triangle}
\end{figure}

However, since we have obtained the full momentum dependence, it is interesting to go
beyond this specific limit. Actually the three-point correlator does not depend
on the three full vectors $\vc{k}_1, \vc{k}_2, \vc{k}_3$, but only on three
scalar quantities, which can be taken to be the three lengths $k_1, k_2, k_3$
(physically this corresponds to statistical isotropy). We can
redefine variables to get the overall magnitude $k \equiv k_1 + k_2 + k_3$ and
two ratios $\gg$ and $\gb$,
\beq\label{plotvars}
\gg \equiv 2 \, \frac{k_2 - k_3}{k},
\qquad
\gb \equiv - \sqrt{3} \, \frac{k_1 - k_2 - k_3}{k},
\eeq
which means that
\beq
k_1 = \frac{k}{2} \lh 1 - \frac{\gb}{\sqrt{3}} \rh,
\qquad
k_{2,3} = \frac{k}{4} \lh 1 \pm \gg + \frac{\gb}{\sqrt{3}} \rh.
\eeq
In addition, because $\vc{k}_1+\vc{k}_2+\vc{k}_3=\vc{0}$, one can use relations
like $|\vc{k}_1+\vc{k}_2|^2 = k_3^2$ and 
$\vc{k}_1\cdot\vc{k}_2 = (k_3^2-k_1^2-k_2^2)/2$.
The domain of $\gg$ and $\gb$ is an equilateral triangle as shown in 
figure~\ref{triangleB}. The vertices of the triangle correspond to one
of the three momenta being zero (the limit for (\ref{three point 2})), 
while the sides correspond to one of the
momenta being equal to half the total sum ($k/2$). From its vertex to the
opposite side $k_s$ grows linearly. Plotting the three-point correlator in such
a way demonstrates its symmetry most clearly, as shown in 
figure~\ref{triangleA}. 
The three-point correlator in that plot has been multiplied by
$k_1^3 k_2^3 k_3^3/((k_1^2+k_2^2+k_3^2)/2)^{3/2}$ to remove the overall $k$
magnitude, and the factor of $(\gk^4/16)(H^4/\tge^2)(2\tge+\tget)$ has also been
omitted (assuming that the momentum dependence of $H_\cH$ and the 
slow-roll parameters can be neglected). There is some dependence on
the relative magnitude of the momenta, but it is only a variation of order unity
for the single-field case.

We finish this section by remarking on the decaying mode. Even though it
disappears very quickly after horizon crossing, one might think that including
it could improve the accuracy of the result, since the window function does 
operate near horizon crossing. When taking into account the decaying mode, 
$\cJ_i$ is no longer
zero, and we have to include the $\gth_i$ equation. Fortunately we can still
solve this system analytically. In the limit of $R \rightarrow 0$ we find that
$\gz^{(1)}$ is unchanged, while $\gz^{(2)}$ picks up an additional term:
$(2\tge_{\cH'}+\tget_{\cH'}) k^2/(k^2+{k'}^2) \, \ga(\vc{k}')$
in (\ref{second sol}) is replaced by
\beq
(2\tge_{\cH'}+\tget_{\cH'}) \frac{k^2}{k^2+{k'}^2} \, \ga(\vc{k}')
+ 3 \sqrt{\frac{\pi}{2}} \frac{1}{c^3} \frac{k^2{k'}^3}
{(k^2+{k'}^2)^\frac{5}{2}} \, \mathrm{i}\ga(\vc{k}').
\eeq
We see that the additional term is not suppressed by slow-roll factors, but by a
factor $1/c^3$ and thus depends on the smoothing length. However, once we
compute the three-point correlator, we find that the additional term drops out
exactly, because of the relative phase factor $\mathrm{i}$. Hence for the final 
result dropping the decaying modes is not an approximation.

{\it Discussion --- }
We have illustrated a new method for calculating non-linearity in 
inflation, which was introduced in \cite{gp2,formalism}, with an explicit
calculation in the single-field case. Analytic resuls were derived to 
second order in a perturbative expansion and to leading order in slow roll.  
We derived a general expression for the three-point correlator of the 
curvature perturbation in Fourier space (the bispectrum). 
In the limit of one of the momenta being much smaller than the other two, 
the expression reduces to a simple result: the scalar spectral index times 
the square of the power spectrum. This agrees with previous results in the 
literature but it was derived in a much simpler way. In particular, it 
agrees exactly with the tree-level action calculation of Maldacena \cite{mald}.
We also demonstrated that the decaying mode makes no contribution to the
three-point correlator at this order.

We introduced convenient variables to plot the full momentum
dependence of the three-point correlator. In \cite{mald} the three-point 
correlator is also given in the limit of all three momenta being equal, which
corresponds with the centre of the triangle in figure~\ref{triangle}. Although 
similar in magnitude, our expression differs in the exact value there. This is 
probably due to our use of a linear stochastic term to emulate the effects of 
sub-horizon perturbations, which does not fully include cubic interactions 
around horizon crossing.  
As is shown in \cite{mf}, the terms causing this discrepancy are 
subdominant in the multiple-field case.

For the single-field case non-Gaussianity is suppressed by slow-roll factors,
and hence unobservable for models that satisfy the CMB constraints. 
However, this is not necessarily the case for multiple-field inflation models 
\cite{uzan,mf}, where the non-Gaussianity can be large. 
The main strength of our method, apart from being simple, 
is that it is easily applicable to general multiple-field inflation; indeed,
the basic equations for the multiple-field case were already presented in
\cite{gp2,formalism}. The formalism is also well-suited for numerical
implementation; no slow-roll approximation is needed, and the end 
result is a real-space realisation, which contains more information than just 
the $N$-point correlator. Results from our numerical implementation of the
formalism will be presented in another paper.
We believe that the simplicity of this formalism, its
general applicability, and its suitability for numerical simulations make it 
very useful for future studies of non-Gaussianity from inflation.

\end{document}